 \documentclass[final,3p,times,twocolumn]{elsarticle}
 \usepackage{graphicx}
 \usepackage{epsfig}

\usepackage{amssymb}
\usepackage{amsmath}

\begin{document}

\begin{frontmatter}



\title{Universal local symmetries in classical mechanics and physical degrees of freedom}


\author[2]{E. Cattaruzza}
\author[1,2]{E. Gozzi}
\address[1]{ Dipartimento di Fisica (sede di Miramare),
Universit\`a di Trieste, Strada Costiera 11, 34151 Trieste, Italy.}
\address[2]{Istituto Nazionale di Fisica Nucleare, Sezione di Trieste, Italy.}

\begin{abstract}
In a recent paper  we have analyzed the role that a universal set of {\it local} symmetries plays in suppressing the superposition principle in classical mechanics via a path integral formulation 
of classical mechanics itself. In this paper first we generalize those local symmetries, second we study the role which the gauge fixing plays and 
 third we put forward the idea of which ones should be the {\it physical} degrees of freedom of the theory.\end{abstract}

\begin{keyword}
Path-Integral Method \sep Classical Mechanics

\end{keyword}

\end{frontmatter}

% \linenumbers

\begin{center}
\begin{figure}[h]
\includegraphics[scale=.65]{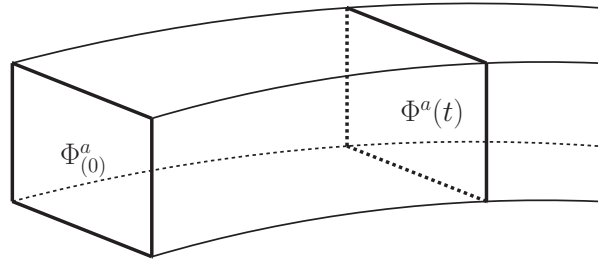}
\caption{\label{FIG1}Block of phase-space in its evolution.}
\vspace{-0.5cm}
\end{figure}
\end{center}

\noindent 1. INTRODUCTION
\vskip 0.3cm
In the 30's Koopman and von Neumann \cite{Koopman,Neumann} (KvN) introduced an operatorial formulation of {\it classical} mechanics (CM). In the late '80 a path integral counterpart of the KvN was developed and the details can be found in ref.\cite{gozzireuter,geomdeq}. We shall indicate this path integral with the acronym CPI (for Classical Path Integral because it describes a {\it classical} system) to distinguish it from the well known Quantum Path Integral of Feynman for which we will use the acronym QPI and which describes a {\it quantum} system.
This last one has a generating functional of the form:
\begin{equation}
\mathcal Z_{QPI}=\int \mathcal D \varphi^a\,e^{\textstyle \frac{i}{\hbar}\int L(\varphi,\dot\varphi)\,dt}\label{QPIGeneratingFunctional}
\end{equation}
where $\varphi^a\equiv(q^1,\dots, q^n;p^1,\dots,p^n),\,a=1,\dots 2\,n$. Unlike the QPI, the CPI  needs the introduction, besides the $\varphi^a$, of $6\,n$ auxiliary variables $c^a,\bar c_a,\lambda_a$ (with $c^a,\bar c_a$ grassmannian) and of two grassmannian partners of time $t$ which we indicate with $\theta,\bar{\theta}$. All these 
variables have an interesting  geometrical interpretation studied in \cite{gozzireuter,regini,gozzimauro}. The whole $8n$ variables $\varphi^a,c^a,\bar c_a,\lambda_a$ can be put together in a multiplet (known in supersymmetric jargon as superfield):
\begin{equation} 
 \Phi^a(t,\theta,\bar{\theta})\equiv \varphi^a+\theta\,c^a+\bar{\theta}\,\omega^{ab}\,\bar{c}_b+i\,\bar{\theta}\theta\,\omega^{ab}\lambda_b\label{SuperPhi},
\end{equation}
where $\omega^{ab}$ is the symplectic matrix \cite{mardsen}. Using $\Phi^a$, the generating functional of the CPI assumes the form \cite{geomdeq}
\begin{equation}
 \mathcal Z_{CPI}=\int \mathcal D\Phi^a
 e^{\textstyle i\int i\,dt\,d\theta\,d\bar{\theta}\,L[\Phi,\dot{\Phi}]} \label{SuperPhiGeneratingFunctional}.
 \end{equation}
Note its formal similarity with the QPI of eqn.(\ref{QPIGeneratingFunctional}). The Lagrangian is the same in both cases and only their argument is changed. Of course in (\ref{SuperPhiGeneratingFunctional}) we do not have $\hbar$ since we are doing classical mechanics. It was shown in ref.\cite{gozzireuter} that $\mathcal Z_{CPI}$ has several {\it global} symmetries. These are due to the presence of the $6n$ auxiliary variables $c^a,\bar c_a,\lambda_a$ which are usually not used in classical mechanics where only $\varphi^a$ is really needed. In \cite{gozzipagani} it was found that (\ref{SuperPhiGeneratingFunctional}) has also some {\it local} invariances. These last symmetries leave $\Phi^a,\,\dot{\Phi}^a$, and as a consequence also $L[\Phi,\dot{\Phi}]$, invariant. The simplest form of these invariances is:   
\begin{align}
\left\{\begin{array}{cc} \label{GPTransformation}
\varphi^a{'}&=\varphi^a+\varepsilon_1(t)\,\theta \,c^a+ \varepsilon_2(t)\,\bar\theta\omega^{ab} \,\bar c^b+i\,\varepsilon_3(t)\bar \theta\theta\,\varphi^a\,\,\,\,\quad \\  
c^a{'}&=c^a-\varepsilon_1(t)\,c^a\hspace{4.9cm}\\
\bar c_a{'}&=\bar c_a\,-\varepsilon_2(t)\,\bar c_a\hspace{4.9cm}\\
\lambda_a{'}&=\lambda_a-\varepsilon_3(t)\,\omega_{ab}\,\varphi^b\hspace{4.3cm}
 \end{array}
 \right.  
\end{align}where $\varepsilon_i(t),\,i=1,2,3$ are three infinitesimal parameters. 
These symmetries are  local because the parameters $\varepsilon(t)$ depend on $t$ and there is also the presence of the partners $\theta,\bar \theta$ of $t$. It was  proved in \cite{gozzipagani} that this local symmetry triggers a superselection 
mechanism which explains why there is no superposition principle in CM. So this symmetry seems to play a crucial role in physics and we thought it's worth to study it further in this paper. We will show that this local invariance helps identifying the true {\it physical} degrees of freedom of a {\it classical} system. This we feel is as important as the derivation of the non-superposition principle in CM which  was obtained in \cite{gozzipagani}.\newline
\noindent 2. GENERALIZED LOCAL SYMMETRIES.
\vskip 0.3cm
It is  easy to show that the superfield $\Phi^a$ and its time derivative are  actually invariant under a more general kind of transformation than the one in eqn.(\ref{GPTransformation}). The most general one is:
\begin{align}
\left\{\begin{array}{cc} 
\varphi^a{'}&=\varphi^a-\theta\,\delta\chi_1^a-\bar \theta\,\delta\chi_5^a+\bar \theta \theta\left[\delta \chi _3^a-\delta \chi _6^a-\delta \chi _9^a\right]\vspace{0.15cm}\\ 
c^a{'}&=c^a+\delta\chi_1^a+\theta\,\delta\chi_2^a+\bar\theta \,\delta\chi_3^a+\bar \theta \theta\, \delta \chi_4^a\hspace{1.4cm}\vspace{0.15cm}\\ 
\bar c^a{'}&=\bar c^a+\delta\chi_5^a+\theta \,\delta \chi_6^a+\bar \theta \,\delta \chi_7^a+\bar \theta \theta \, \delta \chi_8^a\hspace{1.4cm}\vspace{0.15cm}\\
\lambda^a{'}&=\lambda^a+\delta\chi_9^a+\theta \,\delta \chi_{10}^a+\bar \theta \,\delta \chi_{11}^a+\bar \theta \theta \, \delta \chi_{12}^a\hspace{1cm}\vspace{0.cm}
 \end{array} \right.  
 \label{GaugeTransformation}
\end{align}
where $\bar c^a\equiv\omega^{ab}\bar c_b$, $\lambda^a\equiv\omega^{ab} \lambda_b$ and $\delta \chi_i^a,\,i=1,\dots, 12$ are $12$ {\it arbitrary} functions depending each on 12 infinitesimal parameters $\varepsilon_i(t)$ and on the $8n$ variables ($\varphi^a,c^a\,\bar c_a,\lambda_a$).
The 7 functions $\delta \chi_{2}^a,\,\delta \chi_{4}^a,\,\delta \chi_{7}^a,\,\delta \chi_{8}^a,\,\delta \chi_{10}^a,\,\delta \chi_{11}^a,\,\delta \chi_{12}^a$
generate symmetries which are somehow trivial and only due to the grassmannian character of $\theta$ and $\bar \theta$, i.e $\theta^2=\bar \theta^2$. Instead the
remaining 5 functions generate non-trivial symmetries which turn the various components $(\varphi^a,\,c^a,\bar c_a, \lambda_a)$ of the superfield into each other leaving the superfield $\Phi^a$ and its derivative invariant. In order to convince the reader of this fact, let us check this explicitly for one of the 7 functions like ($\delta\chi_2$) and for one of the 5 functions like ($\delta \chi_1$).  Let us start with a  $\delta\chi_2$  transformation:
\begin{equation*}
\delta \varphi^a =  \delta \bar c_a = \delta \lambda_a = 0,\,\,\delta c^a=\theta \,\delta \chi_2^a,
\end{equation*}
so
\begin{align*}
 \delta\Phi^a&= \delta\varphi^a+\theta\,\delta c^a+\bar{\theta}\,\omega^{ab}\,\delta\bar{c}_b+i\,\bar{\theta}\theta\,\omega^{ab}\delta\lambda_b\\
 &=\theta^2\,\delta\chi_2^a=0.
\end{align*}
This is zero  because $\theta^2=0$. We can consider it a sort of "trivial" symmetry because it is only due to the grassmannian character of $\theta$.
Now let us try $\delta \chi_1$:
\begin{equation*}
\delta \varphi^a = -\theta\delta\chi^a_1,\,\,\delta c^a=\delta \chi^a_1,\,\delta \bar c_a = \delta \lambda_a = 0
\end{equation*}
so 
\begin{align*}
 \delta\Phi^a&= \delta\varphi^a+\theta\,\delta c^a+\bar{\theta}\,\omega^{ab}\,\delta\bar{c}_b+i\,\bar{\theta}\theta\,\omega^{ab}\delta\lambda_b\\
 &=-\theta\delta\chi_1^a+\theta\delta\chi_1^a=0.
\end{align*}
In this case the cancellation is not due to the grassmannian character  of $\theta$ or $\bar \theta$ but to a turning of some variables into some others and their mutual cancellations like in standard symmetries. \newline
The careful reader may have  realized that we could  make the choice that the $\delta \chi_i$ depend on more (or less, even zero) parameters $\varepsilon^i(t)$ and not just 12. The invariance of $\Phi$ depends only on which combination of $\delta \chi_i$ enter each transformation in (\ref{GaugeTransformation}). What we mean is that for example $\delta \chi_1$ enters only in $\varphi$ and $c$ and with some particular coefficients in front, so it does not matter on how many parameters $\delta \chi_1$ depends on. The choice of having only 12 parameters is an {\it ansatz} that we adopt here. What cannot  be changed is the maximum number of functions $\delta \chi_i$, we can use. These are at most 12.
The reader may wonder why we have the freedom of 12 functions. Where does this number 12 comes from? The reason is simple, let us rewrite the superfield in this way:
\begin{equation*}
  \begin{array}{c} \Phi^a=(1,\theta,\bar\theta,\bar \theta \theta)\\ \\ \\ \\ \end{array}\hspace{-0.2cm}\left(\begin{array}{c} 
   \varphi^a\\
   c^a\\
   \bar c^a\\
   \lambda^a
   \end{array} \right)
      \end{equation*} 
where in the last step we have introduced a sort of scalar product between a row and a column. 
If we now look  for a transformation which changes the $(\varphi^a,c^a,\bar c^a,\lambda^a)$ but leaves the superfield invariant, we will have: 
  \begin{equation}
  \begin{array}{c} \Phi^a=(1,\theta,\bar\theta,\bar \theta \theta)\\ \\ \\ \\ \end{array}\hspace{-0.2cm}\left(\begin{array}{c} 
   \varphi^a\\
   c^a\\
   \bar c^a\\
   \lambda^a
   \end{array} \right)  \begin{array}{c} =(1,\theta,\bar\theta,\bar \theta \theta)\\ \\ \\ \\ \end{array}\hspace{-0.2cm}\left(\begin{array}{c} 
   \varphi^a{'}\\
   c^a{'}\\
   \bar c^a{'}\\
   \lambda^a{'}
   \end{array} \right).
	\label{SuperPhiInvariace}
      \end{equation}
Setting  
  \begin{equation*}
 \left(\begin{array}{c} 
   \varphi^a{'}\\
   c^a{'}\\
   \bar c^a{'}\\
   \lambda^a{'}
   \end{array} \right)=\underbrace{
  \left(\begin{array}{cccc}
   \alpha_1&\beta_1&\gamma_1&\delta_1  \\
   \alpha_2&\beta_2&\gamma_2&\delta_2  \\
   \alpha_3&\beta_3&\gamma_3&\delta_3  \\
   \alpha_4&\beta_4&\gamma_4&\delta_4  
   \end{array}\right)}_{ \displaystyle  \mathbb A}
   \hspace{0.1cm}\left(\begin{array}{c} 
   \varphi^a\\
   c^a\\
   \bar c^a\\
   \lambda^a
   \end{array} \right),
      \end{equation*}
where the $4\times 4$ matrix $\mathbb A$ is made of 16 functions which depend  on $\theta,\bar \theta, \varphi^a,c^a,\bar c^a,\lambda^a$
and on the infinitesimal parameter $\varepsilon_i(t)$, we get from (\ref{SuperPhiInvariace}) that:
\begin{equation}
(1,\theta,\bar\theta,\bar \theta \theta)\cdot \mathbb A= (1,\theta,\bar\theta,\bar \theta \theta),\label{PhiConstrain}
\end{equation}
where  "$\cdot$" is the usual row-column product.
The  relation (\ref{PhiConstrain}) above defines 4 constraints. So the 16 functions we have in the matrix $\mathbb A$ are reduced effectively to 12  free functions like we have in the transformation (\ref{GaugeTransformation}).\\
\vskip 0.2cm
\noindent 3. EQUATIONS OF MOTION.
\vskip 0.3cm
Next,  we shall explore in which sense these transformations are symmetries of the system. As we said they leave invariant  the super field $\Phi$ and its time derivative so they are invariances of $L[\Phi]$ in (\ref{SuperPhiGeneratingFunctional}). The equations of motion \cite{gozzireuter} for $\Phi$ are:
\begin{equation}
\dot \Phi^a=\omega^{ab}\frac{\partial H[\Phi]}{\partial \Phi^b} \label{SuperPhiEQM}
\end{equation}
where $H$ is the Hamiltonian associated to $L$, but with $\varphi^a$ replaced by $\Phi^a$.\newline 
Because both $\Phi^a$ and $\dot\Phi^a$ are invariant, clearly also the Hamiltonian equations above have the same property.
Expanding (\ref{SuperPhiEQM}) in $\theta,\bar \theta$ we get:  
\begin{align}
&\left(\dot\varphi^a-\omega^{ab}\partial_bH\right)+\theta\left(\dot c^a-\omega^{ab}\partial_{b}\partial_{k}H\,c^k\right)\label{SuperPhiComponents}\\
&+\bar\theta\left(\dot{\bar{c}}^a-\omega^{ab}\partial_{b}\partial_{k}H\bar c^k\right)\nonumber\\&+\bar\theta \theta\left(\dot \lambda^a-\omega^{ab}\partial_{b}\partial_{k}H\lambda^k+\omega^{ab}\partial_{b}\partial_{l}\partial_{m}H\,\bar c^l\,c^m\right)=0.
\nonumber
\end{align}
In order that the whole  expression above  be zero, each term multiplying $1,\theta,\bar \theta,\bar \theta \theta$ must be zero and these are exactly the equations of motion of respectively $\varphi^a,c^a,\bar c^a,\lambda^a$. An obvious question to ask is if the single equations of $\varphi^a,c^a,\bar c^a,\lambda^a$ are 
invariant under the  local transformations (\ref{GaugeTransformation}). The answer is no! 
For example doing a simple transformation of the type (\ref{GaugeTransformation}) with  only $\delta \chi_1^a\neq 0$, we 
get for the variation of the equation  of motion of $\varphi,c,\bar c,\lambda$: 
\begin{align}
&\delta\left(\dot\varphi^a-\omega^{ab}\partial_bH\right)=\theta\left(-\delta\dot{\chi}_1^a+\omega^{ab}\partial_b\partial_kH\,\delta\chi_1^k\right)
\nonumber\\
&\delta\left(\dot c^a-\omega^{ab}\partial_{b}\partial_{k}H\,c^k\right)=\left(\delta\dot\chi_1^a-\omega^{ab}\partial_b\partial_kH\,\delta\chi_1^k\right)\nonumber\\
&\hspace{3.3cm} +\theta\left(\omega^{ab}\partial_b\partial_k\partial_lH\,\delta\chi_1^l\,c^k\right)\nonumber\\
&\delta\left(\dot{\bar{c}}^a-\omega^{ab}\partial_{b}\partial_{k}H\bar c^k\right)\,=\theta\left(\omega^{ab}\partial_b\partial_k\partial_lH\,\delta\chi_1^l\,\bar c^k\right)\label{PhiVariation}\\
&\delta\left(\dot \lambda^a-\omega^{ab}\partial_{b}\partial_{k}H\lambda^k+\omega^{ab}\partial_{b}\partial_{l}\partial_{m}H\,\bar c^l\,c^m\right)=\nonumber\\
&\theta\left(\omega^{ab}\partial_b\partial_k\partial_lH\,\delta\chi_1^l\,\lambda^k-\omega^{ab}\partial_b\partial_k\partial_l\partial_mH\,\delta\chi_1^m\bar c^kc^l\right)\nonumber\\
&\hspace{3.3cm}+\omega^{ab}\partial_b\partial_k\partial_lH\,\bar c^k\delta\chi_1^l.\nonumber
\end{align}
The various pieces which  appear on the RHS. of (\ref{PhiVariation}) must combine to give a total zero variation in (\ref{SuperPhiComponents}) confirming that eq.(\ref{SuperPhiEQM}) for the superfield is invariant.
So while $\Phi$ is a gauge (i.e {\it local}) invariant quantity, the components ($\varphi^a,c^a,\bar c^a,\lambda^a$) and their equations are not.
An intuitive understanding of this can be obtained via the following picture. The equation of motion for the superfield (\ref{SuperPhiEQM}) 
can be solved by giving an initial superfield configuration $\Phi_0^a$. Let us represent the superfield via a square. Its time evolution  
can be represented as the tube in Figure.\ref{FIG1}.  Giving an initial $\Phi_0^a$ this does not fix the initial components $\varphi^a_{(0)},c^a_{(0)},\bar c^a_{(0)},\lambda^a_{(0)}$ in a unique manner.
 In fact there are many initial conditions on the components  which give the same $\Phi^a_0$. For example all  those related by the transformation (\ref{GPTransformation}) and (\ref{GaugeTransformation}) give the same $\Phi_0^a$ because the transformation (\ref{GPTransformation}) and (\ref{GaugeTransformation}) leave the superfield invariant. As these initial conditions for $\varphi,c,\bar c,\lambda$ are different from each other also their trajectories will be different as indicated in Figure \ref{FIG2}. The points on the initial square correspond to the different initial conditions $\varphi^a_{(0)}$. From here one gathers that $\Phi^{a}$ is "like" a cell of phase-space.The fact that the equation of motion of $\varphi^a$ is not gauge invariant is signaled here by the various trajectories which $\varphi^a$ can have in correspondence to a unique trajectory for $\Phi^a$. We will provide now a different analysis, more mathematically founded, which will reach the same conclusion. When an equation of motion is not gauge invariant a way out is to add  gauge-fields. Let us do that for the equation of motion (EM) of $\varphi^a,c^a,\bar c^a,\lambda^a$, 
which we will indicate in a compact way as $(\textrm{EM})^{\varphi},(\textrm{EM})^{c},(\textrm{EM})^{\bar c},(\textrm{EM})^{\lambda}$. Let us add gauge fields so that the new equations of motion are:
\begin{align}
&\begin{array}{cc} 
&(\textrm{EM})^{\varphi}+A_1^{\varphi} +\theta \,A_{\theta}^{\varphi}+\bar \theta \,A_{\bar \theta}^{\varphi}+\bar \theta \theta \,A_{\bar \theta \theta}^{\varphi}=0\,\vspace{0.1cm}\\  
&(\textrm{EM})^{c}+A_1^{c} +\theta \,A_{\theta}^{c}+\bar \theta \,A_{\bar \theta}^{c}+\bar \theta \theta \,A_{\bar \theta \theta}^{c}=0\,\,\,\,\vspace{0.1cm}\\  
&(\textrm{EM})^{\bar c}+A_1^{\bar c} +\theta \,A_{\theta}^{\bar c}+\bar \theta \,A_{\bar \theta}^{\bar c}+\bar \theta \theta \,A_{\bar \theta \theta}^{\bar c}=0\,\,\,\,\vspace{0.1cm}\\  
&(\textrm{EM})^{\lambda}+A_1^{\lambda} +\theta \,A_{\theta}^{\lambda}+\bar \theta \,A_{\bar \theta}^{\lambda}+\bar \theta \theta \,A_{\bar \theta \theta}^{\lambda}=0.\,\,\\  
 \end{array}  
\label{EMGauge}
\end{align}
\begin{center}
\begin{figure}
\includegraphics[scale=.65]{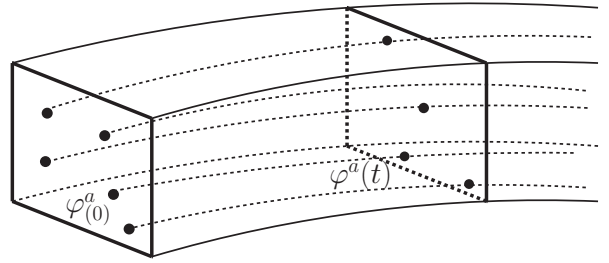}
\caption{\label{FIG2} Block of phase-space and the trajectories of its components.}
\end{figure}
\vspace{-1.cm}
\end{center}
\vskip 0.2cm
We have tried other manners to insert the gauge fields where a more direct coupling between the gauge-fields and the "matter" fields $\varphi,c,\bar c,\lambda$ was present , but they are more difficult to handle especially regarding the issue of the  gauge-fixing.  
All together in eqn.(\ref{EMGauge}) we have 16 gauge fields but we know that when we sum up the equations (\ref{EMGauge}), like we did in (\ref{SuperPhiComponents}),
we have to obtain the eqn.(\ref{SuperPhiEQM}) for the superfield  which does not contain any gauge field.  This canceling of the gauge fields leads to 4 constraints among them, and this brings the total number down from 16 to 12,  exactly as many as the arbitrary functions present in the transformations (\ref{GaugeTransformation}). Because of these constraints, eqn.(\ref{EMGauge}) can be reduced to the form:
\begin{align}
\begin{array}{cc} 
(\textrm{EM})^{\varphi}-\theta\,A_1^{c}-\bar \theta \,A_{1}^{\bar c}+\bar \theta \theta\left(A_{\bar \theta}^{c}-A_{\theta}^{\bar c}-A_{1}^{\lambda}\right)=0\,\vspace{0.1cm}\\  
(\textrm{EM})^{c}+A_1^{c} +\theta \,A_{\theta}^{c}+\bar \theta \,A_{\bar \theta}^{c}+\bar \theta \theta \,A_{\bar \theta \theta}^{c}=0\quad\quad\quad\,\vspace{0.1cm}\\  
(\textrm{EM})^{\bar c}+A_1^{\bar c} +\theta \,A_{\theta}^{\bar c}+\bar \theta \,A_{\bar \theta}^{\bar c}+\bar \theta \theta \,A_{\bar \theta \theta}^{\bar c}=0\quad\quad\quad\,\vspace{0.1cm}\\  
(\textrm{EM})^{\lambda}+A_1^{\lambda} +\theta \,A_{\theta}^{\lambda}+\bar \theta \,A_{\bar \theta}^{\lambda}+\bar \theta \theta \,A_{\bar \theta \theta}^{\lambda}=0.\quad\quad\,\,\,\,\,\\  
 \end{array}  
\label{InvariantEQGauge}
\end{align}
Here for simplicity we have suppressed the index "$a$" which the gauge fields should have like the variables ($\varphi^a,c^a,\bar c^a,\lambda^a$) have.   
The reader must have realized  that the first equation in (\ref{InvariantEQGauge}) does not have the usual  form given by Hamilton to the equation for $\varphi^a$  but contains some extra pieces due to the gauge-fields.  The way out is to prove that we can choose a gauge fixing which brings those fields to zero. If this  can be achieved then we can say that the usual Hamilton equations are nothing else that the gauge fixed version of more general equations. Let us start by finding how  the gauge fields must transform in order for the eq.(\ref{InvariantEQGauge}) to be invariant. Let us limit ourselves to a $\delta\chi_1^a$ transformation in (\ref{GaugeTransformation}). After some  calculations we get that the gauge fields appearing in (\ref{InvariantEQGauge}) must transform as follows:
\begin{equation}
\left\{\begin{array}{ccc} 
(\delta A_1^c)^a&=&-\delta\dot{\chi}_1^a+\omega^{ab}(\partial_b\partial_kH)\,\delta\chi_1^k\quad\vspace{0.15cm}\\ 
(\delta A_{\theta}^c)^a&=&-\omega^{ab}(\partial_b\partial_k\partial_l H)\,\delta\chi_1^l\,c^k\quad\quad\vspace{0.15cm}\\ 
(\delta A_{\theta}^{\bar c})^a&=&-\omega^{ab}(\partial_b\partial_k\partial_l H)\,\delta\chi_1^l\,\bar c^k\quad\quad\vspace{0.15cm}\\ 
(\delta A_1^{\lambda})^a&=&-(\delta A_{\theta}^c)^a\hspace{2.8cm}\vspace{0.15cm}\\ 
(\delta A_{\theta}^{\lambda})^a&=&-\omega^{ab}(\partial_b\partial_k\partial_l H)\,\delta\chi_1^l\,\lambda^k+\quad\,\vspace{0.15cm}\\ 
&+&\omega^{ab}
(\partial_b\partial_k\partial_l\partial_m H)\,\delta\chi_1^m\,\bar c^k\,c^l.
 \end{array} \right.  
 \label{GaugeFieldTransformation}
\end{equation}
The other gauge-fields do not change under a $\delta \chi_1$ transformation.
So the $\delta\chi_1$ transformation involves only 5 of the 12 gauge fields.  Consequently let us make the following ansatz  for  $\delta \chi_1$:
\begin{equation} 
\delta\chi_1^a\equiv \sum_{\alpha =1 }^5
\varepsilon^{\alpha }(t)\,\chi_{1\alpha}^a\end{equation}
where each infinitesimal parameter $\varepsilon^{\alpha }$ corresponds to a different gauge field. We have neglected, for simplicity of notation, an index "$a$" on the infinitesimal parameter like we neglected it on the gauge fields. If we want to bring the gauge fields to zero, we should perform a gauge transformation such that:  
\begin{equation}
\left\{\begin{array}{ccc} 
A_1^c{'}=0\Longrightarrow \delta A_{1}^c=-A_1^c\vspace{0.15cm}\\ 
A_{\theta}^c{'}=0\Longrightarrow \delta A_{\theta}^c=-A_{\theta}^c\vspace{0.15cm}\\ 
A_{\theta}^{\bar c}{'}=0\Longrightarrow \delta A_{\theta}^{\bar c}=-A_{\theta}^{\bar c}\vspace{0.15cm}\\ 
A_1^{\lambda}{'}=0\Longrightarrow \delta A_{1}^{\lambda}=-A_1^{\lambda}\vspace{0.15cm}\\ 
A_{\theta}^{\lambda}{'}=0\Longrightarrow \delta A_{\theta}^{\lambda}=-A_{\theta}^{\lambda}.\vspace{0.15cm}\\ 
 \end{array} \right.  
 \label{GFC}
\end{equation}
Putting together (\ref{GFC}) and (\ref{GaugeFieldTransformation}) we get 
\begin{equation}
\left\{\begin{array}{ccc} 
(- A_1^c)^a&=&-\delta\dot{\chi}_1^a+\omega^{ab}(\partial_b\partial_kH)\,\delta\chi_1^k\quad\vspace{0.15cm}\\ 
(- A_{\theta}^c)^a&=&-\omega^{ab}(\partial_b\partial_k\partial_l H)\,\delta\chi_1^l\,c^k\quad\quad\vspace{0.15cm}\\ 
(- A_{\theta}^{\bar c})^a&=&-\omega^{ab}(\partial_b\partial_k\partial_l H)\,\delta\chi_1^l\,\bar c^k\quad\quad\vspace{0.15cm}\\ 
(- A_1^{\lambda})^a&=&-(\delta A_{\theta}^c)^a\hspace{2.8cm}\vspace{0.15cm}\\ 
(- A_{\theta}^{\lambda})^a&=&-\omega^{ab}(\partial_b\partial_k\partial_l H)\,\delta\chi_1^l\,\lambda^k+\quad\,\vspace{0.15cm}\\ 
&+&\omega^{ab}
(\partial_b\partial_k\partial_l\partial_m H)\,\delta\chi_1^m\,\bar c^k\,c^l.
 \end{array} \right.
 \label{GFCGaugeFieldTransformation}
\end{equation}
Remember that the $\delta\chi_{1}^a$ are functions also of ($\varphi^a,c^a,\bar c^a,\lambda^a$) and for  these variables we will insert their classical trajectories $\varphi^a(t;\varphi_{(0)}^a),c^a(t;c_{(0)}^a),\bar c^a(t;\bar c_{(0)}^a),\lambda^a(t;\lambda_{(0)}^a)$, where we have indicated with $\varphi^a_{(0)},c^a_{(0)},\bar c^a_{(0)},\lambda^a_{(0)}$ the initial conditions. The eqn.(\ref{GFCGaugeFieldTransformation}) becomes then a linear  system  of differential equations that we can solve for the parameter $\varepsilon^{\alpha }(t)$. Of course the final solution will depend on the functional form we choose for $\delta\chi_{1}^a$  and also on the initial conditions $\varphi^a_{(0)},c^a_{(0)},\bar c^a_{(0)},\lambda^a_{(0)}$ given to the trajectories. This last fact tells us that the gauge fixing is somehow linked to the initial conditions $\varphi^a_{(0)}$  as we had explained in our earlier qualitative analysis. 
So given a particular gauge field configuration $A(t)$, the solution of eqn.(\ref{GFCGaugeFieldTransformation}) will provide us with the form of the gauge transformation (i.e the dependance on $t$ of the parameters $\varepsilon(t)$) which brings the gauge fields to zero. Note that there apparently still remain the total freedom in the functional form 
of $\delta\chi_{1}^a$. Actually it is not really so. In fact it is possible to prove that, for the solution of (\ref{GFCGaugeFieldTransformation}) to exist, we have to impose some constraints on the functional forms of $\delta \chi_1^a$. Details are too long to be given here. 
With (\ref{GFCGaugeFieldTransformation}) we have managed to put to zero only 5 of the 12 gauge fields. It is easy to prove that with a $\delta\chi_5$ and $\delta\chi_3$ transformation we can put to zero also the other 7 fields. This, in turn, will impose some constraints on the functional form of $\delta \chi_3^a$ and $\delta \chi_5^a$. So the gauge fixing limits in part  the form of three of the functions $\delta \chi^a_i, i=1,3,5$. An issue which is still open is the physical and mathematical meaning of the total functional freedom we have in the remaining functions.\\
\vskip 0.2cm

\noindent 4. CONCLUSION.
\vskip 0.3cm
We summarize here some of the conclusions we can draw from this work:
\begin{enumerate}
\item the gauge-invariant degrees of freedom in CM are the superfield $\Phi^a$ and not the phase-space point $\varphi^a$.
\item The gauge freedom we have in this formulation is somehow related to the freedom in the choice of the initial conditions of $\varphi^a$.
\item The $\Phi^a$, as they are gauge invariant, should be considered the true {\it physical} degrees of freedom. It is exactly like in gauge theory where the true {\it observables} and {\it physical} variables are the gauge invariant ones.  
\item There is another reason to consider the $\Phi^{a}$ the true {\it physical} degrees of freedom. If QM is the most fundamental theory of nature , then in CM we will be able to predict the motion only of cells of phase-space of volume bigger than $\hbar$. These will be the true degrees of freedom in CM. May it be that these cells are represented by the super fields $\Phi^a$?. We showed before that to the same $\Phi^{a}_{0}$ there corresponds many $\varphi^{a}_{(0)}$ so $\Phi^{a}_{0}$ is really like a block of phase-space. 

\item We can recover the usual Hamilton equations for $\varphi^a$ as a gauge-fixed version of a different set of equations.
\end{enumerate}
One could ask why, if $\Phi^a$ are the true physical degrees of freedom, Hamilton managed to do all the correct physics  using $\varphi^a$ and not $\Phi^a$. The reason is that the two variables are related to each other, as it was proved in ref.\cite{geomdeq}. Moreover, by properly choosing a gauge fixing, they have the same equations of motion as we have shown in this paper. So  somehow   Hamilton was using a particular variable, $\varphi^a$, which was {\it mathematically} related to the true {\it physical} degrees of freedom $\Phi^a$, but it was not gauge invariant. \newline The work we are pursuing now is to start from the QPI and perform a sort of "block-spin" transformations like it is done in renormalization-type procedures. Once these blocks get to be much bigger that $\hbar$ we should end up in
CM in a path-integral form, but this should be  nothing else than the CPI. If this can be achieved then the physical interpretation 
we have given of $\Phi^a$  as blocks of phase-space should be the correct one. If in this "renormalization-like" procedure we get the CPI then, as the Lagrangian remains the same in form as in the QPI , we can say that only the field get "renormalized" from $\varphi^a$ to $\Phi^a$ while the mass and the couplings remain the same. At that point it will be very interesting to study the fixed points of this "renormalization-type" procedure and see to which theories these fixed points  correspond.
We think that this approach   will throw a whole new light on the interface  between QM and CM.
\\
\vskip0.3cm
\noindent{\bf ERRATA CORRIGE}
\vskip 0.3cm
All the part on the gauge-fixing from eq. (11) till the end of section 3 should be disregarded. The reason is that,
even bringing to zero the {\it naive} gauge field we used, this  does not allow the equations
of motion  for $\varphi^a$ to keep  the standard Hamiltonian form . The conclusion are nevertheless safe  and are even stronger because they confirm that only the superfield $\Phi^a$ is the physical gauge invariant degree of freedom. The only link between $\varphi^a$ and  $\Phi^a$ is that they  are related to each other by  change of  "picture" in the grassmannian time, as proved in ref.[4].
\vskip0.5cm
\par{\it  Acknowledgements:} This work has been financially supported by grants from MIUR (PRIN 2008), INFN (GE41) and University of Trieste (FRA-grants). We like to thank M.Reuter and C. Pagani for their support of this project.



%% References


\bibliographystyle{model1-num-names}
\bibliography{<your-bib-database>}



\end{document}